\documentclass[twocolumn,showpacs,preprintnumbers,amsmath,amssymb,superscriptaddress]{revtex4}

\usepackage[draft]{hyperref}
\usepackage{amsmath}

\usepackage{hyperref}%
\usepackage{graphicx}
\usepackage{dcolumn}
\usepackage{bm}
\usepackage{epsf}
\usepackage{mathrsfs}
\usepackage{color}

\begin{document}


\title{Reversible nonreciprocity in photonic structures \\ infiltrated with liquid crystals }

\author{Andrey E. Miroshnichenko}

\affiliation{
Nonlinear Physics Centre and Centre for Ultra-high bandwidth Devices for Optical Systems (CUDOS)\\
Australian National University, Canberra ACT 0200, Australia}

\author{Etienne Brasselet}
\affiliation{Centre de Physique Optique Mol\'eculaire et Hertzienne, Universit\'e Bordeaux 1, CNRS,\\
351 Cours de la Lib\'eration, 33405 Talence Cedex, France}

\author{Yuri S. Kivshar}

\affiliation{
Nonlinear Physics Centre and Centre for Ultra-high bandwidth Devices for Optical Systems (CUDOS)\\
Australian National University, Canberra ACT 0200, Australia}

\begin{abstract}
We demonstrate how to achieve reversible nonreciprocal optical response in a periodic photonic structure with a pair of defects, one of them being a nonlinear liquid crystal defect layer. The twin defect layers structure is symmetric at low intensity and becomes asymmetric above a power threshold corresponding to the optical reordering of the liquid crystal. We show that nonreciprocal effects {\em can be reversed} by changing the wavelength as a consequence of the defect mode dependent light localization inside the structure.
\end{abstract}


\maketitle

Optical nonreciprocity (ONR) is referred to different properties for opposite propagation directions of electromagnetic waves. Nonreciprocal response is usually related to time-reversal symmetry breaking of light-matter interaction~\cite{rjp:rpp:04}. A well-known example can be found in gyrotropic materials, where left- and right-handed polarized fields propagate at different speeds in the presence of an external static magnetic field~\cite{zvezdin,favi:pre:01,khan}. However, ONR is not restricted to the use of magnetic field. Indeed, it has been suggested that time-dependent refractive index modulation can lead to ONR by inducing dynamically indirect interband photonic transitions~\cite{zysf:np:09}. Nonreciprocity can also be achieved without the use of applied external fields in spatially asymmetric light sensitive media~\cite{rjp:rpp:04}, and, formally, one can make the distinction between {\em linear} and {\em nonlinear} systems. The former class is based on optical absorption~\cite{gsasdg:ol:02} or anisotropy~\cite{lc_diode}, or employs optomechanical effects~\cite{smjtrml:prl:09}. On the other hand, the use of optical nonlinearities allows to achieve with device tunability driven by the light itself. For example, the power dependent ONR has been proposed for the realization of all-optical diodes~\cite{sfmysk:josa:02,zkzsxavg:jap:06,pumpassisting_OE2008} and unidirectional couplers~\cite{aaga:ol:08}.

\begin{figure*}[t]
\center{\includegraphics[width=1.7\columnwidth ]{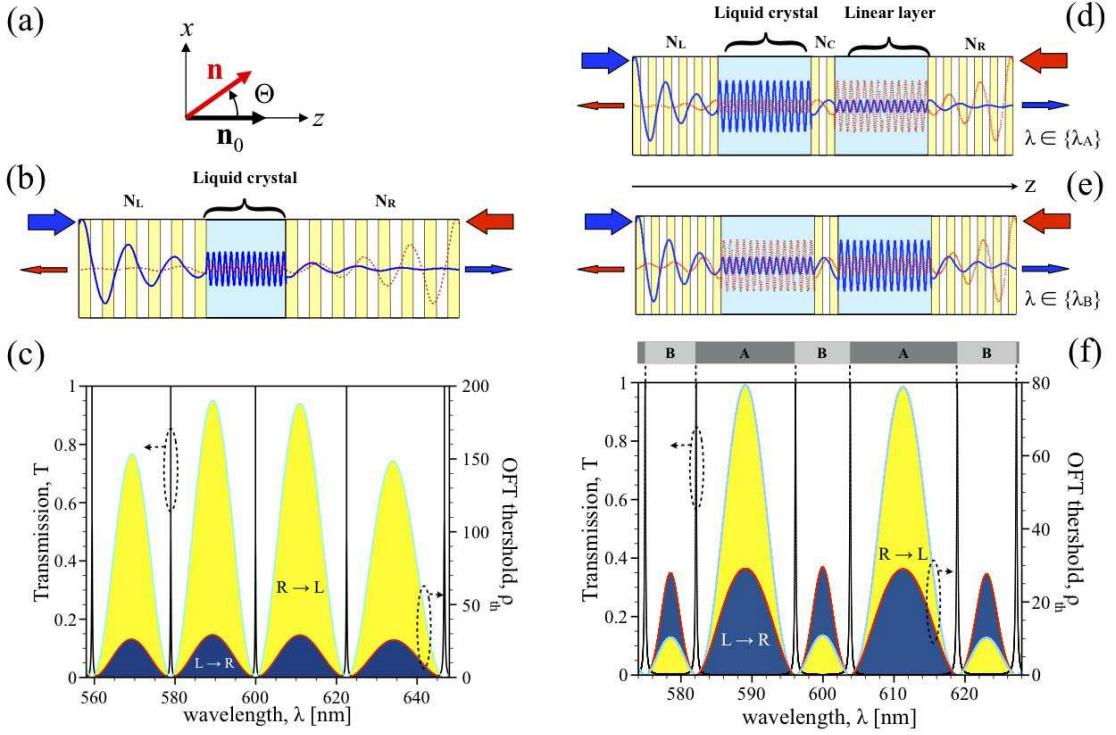}}
\caption{\label{fig:fig1}
(Color online) (a) Director representation characterized by the angle $\Theta(z)$ with respect to the unperturbed state ${\bf n}_0 = {\bf e}_z$. (b) Case I: asymmetric  single LC defect periodic structure with a different number of periodic layers on the left and right sides, $N_L \neq N_R$. Typical electric field amplitude profile for left-to-right light propagation, `LR', (resp. right-to-left, `RL') in the absence of reorientation. (c) Transmission $T$ and normalized threshold intensity $\rho_{\rm th}^{\rm LR, RL}$ with $N_L=5$ and $N_R=7$ for case I. Right part, case II: (d,e) Electric field profile within the unperturbed twin defect layer structure for LR and RL propagation and a wavelength belonging to region referred to as `A' [panel (d)] or `B' [panel (e)], see panel (f). $T$ and $\rho_{\rm th}^{\rm LR, RL}$ with $N_L=N_R=5$ and $N_C=1$.}
\end{figure*}

Among various optically nonlinear materials, liquid crystals (LCs) are known to possess the strongest nonlinear response, which is up to several orders of magnitude larger than Kerr nonlinearity of conventional dielectrics~\cite{ick:pr:09}. Moreover, the efficiency of all-optical nonreciprocal strategies based on LCs can be enhanced in dielectric periodic structures by embedding a LC defect layer asymmetrically inside it. Such a geometry has been previously explored in \cite{aem_OE06}, but for an hypothetical LC material exhibiting the first-order optical Fr\'edericksz transition (OFT) in the absence of periodic structure \cite{tnvsavzy:mclc:86}, and in the particular case $\lambda = \lambda_{\rm d}$ where $\lambda$ is the wavelength of light and $\lambda_{\rm d}$ is a defect mode wavelength. In fact, as shown in the present work, such a nonreciprocal behavior is more general and applies to common nematic LCs, whatever $\lambda$ is. Nevertheless, the corresponding optical diode behavior is intrinsically unidirectional. Here we propose a strategy enabling all-optical diode operation to work both for light propagation toward $\pm z$, thus introducing the concept of all-optical reversible ONR, which is controlled by the wavelength.

As an example, we consider a photonic structure created by alternating layers of SiO$_2$ and TiO$_2$ with thicknesses $d_{\rm \rm SiO_2}=103$nm and $d_{\rm \rm TiO_2}=64$nm, respectively, whose transmission spectrum has a gap in the visible range between $500$ and $720$nm. Inside such a structure either one (case I) or two (case II) defect layers are embedded. In case I, the defect is a nematic layer with its director (i.e. the unit vector ${\bf n}$ that represents the local average molecular orientation of the LC) lying along the light propagation axis $z$ in the absence of reorientation, ${\bf n}_0 = {\bf e}_z$ [see Fig.~1(a)]. Moreover, the numbers of periodic layers differs between the left and right side, $N_L \neq N_R$ [see Fig.~1(b)]. In case II, there is an additional defect linear layer with the same optical path length than the LC defect, but with $N_L = N_R$ [see Fig.~1(d,e)].

The optical response of these two systems under a linearly polarized light is obtained following the standard Berreman's $4\times4$ matrix formalism and taking into account the optical orientational nonlinearities of the LC layer \cite{dwb:josa:72,aemebysk:apl:08,aemebysk:pra:08}. Without loss of generality, the calculations are performed with the refractive indices $n_\perp=1.5$ and $n_\parallel=1.7$, where `$\perp$' and `$\parallel$' refer to a direction perpendicular and parallel to ${\bf n}$, respectively, and the ratio between splay and bend Frank elastic constants is $K_1/K_3=2/3$. We choose defect layer thicknesses $L=5\mu$m for both cases, which means that the linear defect layer in case II has a refractive index $n_\perp$.

\begin{figure}[t]
\includegraphics[width=\columnwidth ]{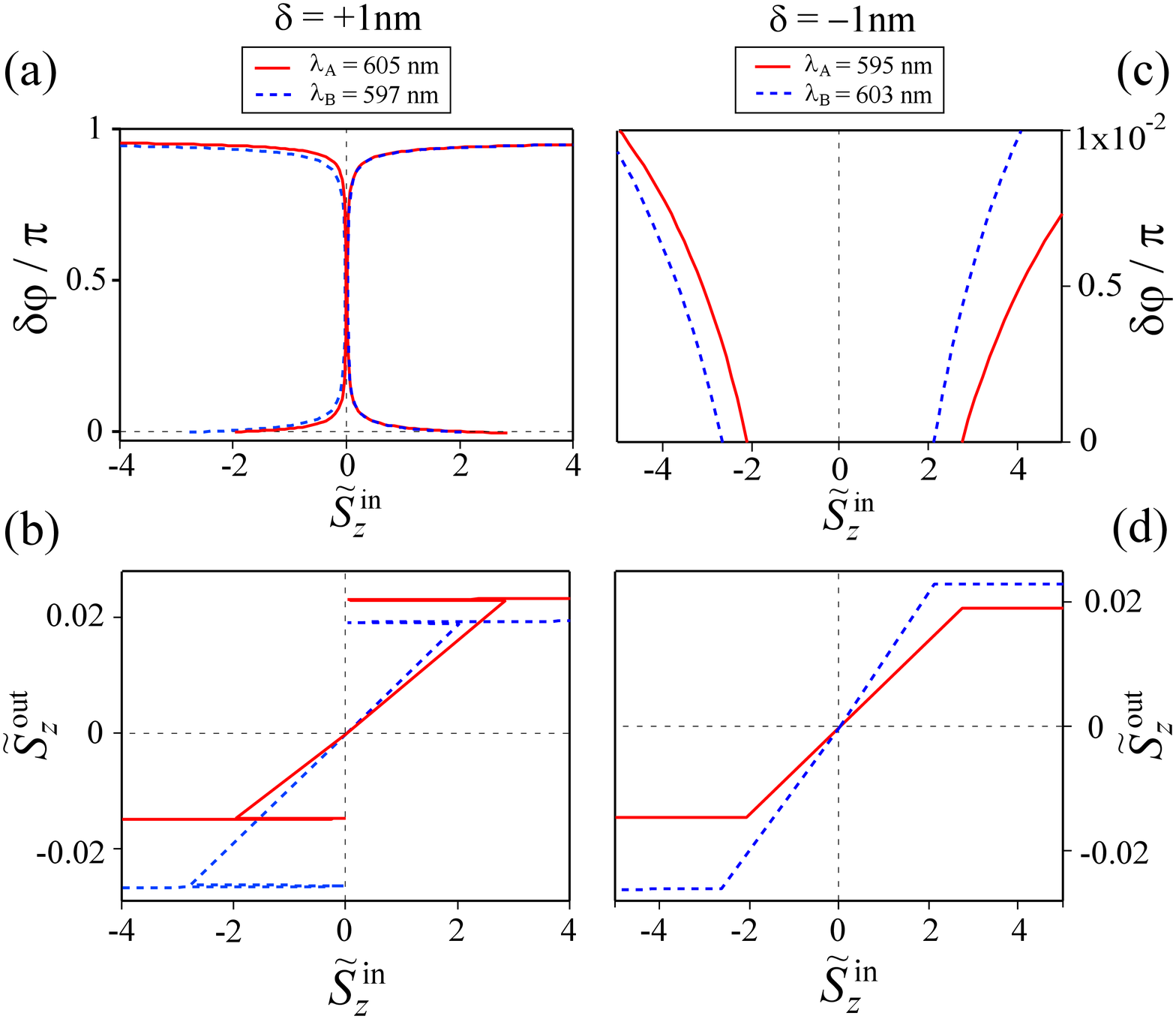}
\caption{\label{fig:fig2} (Color online) Phase jump of the transmitted light field, $\delta \varphi$, and output Poynting vector amplitude normalized to its threshold value, $\widetilde{S}_z^{\rm out} = \widetilde{S}_z^{\rm out}/\widetilde{S}_{z {\rm , th}}^{\rm in}$, vs. $\widetilde{S}_z^{\rm in}$ for positive (a,b) and negative (c,d) values of detuning parameter $\delta = \lambda - \lambda_{\rm d}$ in the vicinity of the two adjacent defect modes $\lambda_{\rm d} \simeq 604$nm and $\lambda_{\rm d} \simeq 596$nm. Solid and dashed curves refer to wavelength belonging to region A and B respectively, see Fig.~1(f).}
\end{figure}

In case I, the field distribution inside the unperturbed nonlinear photonic structure is {\em asymmetric}, as depicted in Fig.~1(b) where a typical calculated electric field profiles are shown for left-to-right (LR) and right-to-left (RL) incident field with wavelength $\lambda$ is shown. Obviously, the light confinement inside the LC layer depends on the propagation direction. Therefore, the light intensity threshold above which optical reordering occurs differs for the LR and RL situations. The corresponding normalized thresholds $\rho_{\rm th}$ are shown in Fig.~1(c) where we introduce the normalized incident intensity with respect to the OFT value for the LC slab alone, $\rho = I_{\rm incident}/I_{\rm OFT}$. As expected, these thresholds are spectrally modulated with local minima at each of the defect mode frequency~\cite{aemebysk:apl:08} and we find that $\rho_{\rm th}^{\rm LR}<\rho_{\rm th}^{\rm RL}$ independently of $\lambda$ [see Fig.~1(c)]. The optical response of the whole structure is thus reciprocal for $\rho<\rho_{\rm th}^{\rm LR}$ and nonreciprocal for $\rho>\rho_{\rm th}^{\rm LR}$ following the generic scheme relying on nonlinearity~\cite{sfmysk:josa:02,zkzsxavg:jap:06,pumpassisting_OE2008}. The unidirectionality of such optical diodes, however, is imposed by the intrinsic asymmetry of the structure that is present even when the nonlinearity is not activated.

From above considerations, to enhance the all-optical control of the unidirectionality thus requires an initially symmetric device. The nonreciprocal condition, however, imposes a nonlinear asymmetry. Such conditions are satisfied in the geometry of the case II, where a nonlinear/linear pair of defects are embedded into a symmetric periodic structure, $N_R=N_L$, and separated by a symmetric multilayer characterized by $N_C$ [see Fig.~1(d,e)]. The LR and RL normalized reorientation thresholds are shown in Fig.~1(f). We find that one obtains either $\rho_{\rm th}^{\rm LR}<\rho_{\rm th}^{\rm RL}$ or $\rho_{\rm th}^{\rm LR}>\rho_{\rm th}^{\rm RL}$ depending on the wavelength, as illustrated by the spectral zones referred to as `A' and `B', respectively [see Fig.~1(f)]. As a matter of fact, such a behavior is consistently retrieved from the spatial confinement of an incident light field into the unperturbed LC defect layer, as shown from the typical electric field profiles calculated for the LR and RL situations in the case A, where confinement is more efficient for LR propagation  [see Fig.~1(d)], and in the case B, where RL confinement is better [see Fig.~1(e)].

As a result, ONR {\em can be reversed by changing the wavelength of light}.  Note, however, that the spatial mirror symmetry for the light field is restored when $\lambda = \lambda_{\rm d}$, which results in identical thresholds $\rho_{\rm th}^{LR}(\lambda_{\rm d})=\rho_{\rm th}^{RL}(\lambda_{\rm d})$. Such a behavior originates from an interplay between even and odd defect modes supported by this structure, hence, the field distribution inside defect layers depends not only on the direction of propagation, but on the input wavelength as well.

When $\min[\rho_{\rm th}^{LR,RL}(\lambda)] < \rho < \max[\rho_{\rm th}^{LR,RL}(\lambda)]$ at a given wavelength, $\Theta(z) \neq 0$, see Fig.~1(a), for light impinging from one direction, whereas $\Theta(z) = 0$ for the opposite light propagation direction. This can be interpreted as `on' and `off' orientational states. Above $\max[\rho_{\rm th}^{LR,RL}(\lambda)]$, the LC is always in the `on' state whatever the incident light propagation is, however, the LR and RL reoriented states remains distinct, thus preserving a nonreciprocal material response. Also, we note that the intensity is not the only parameter that dictates the optically induced reorientation qualitative behavior, indeed, it is known that the detuning parameter $\delta=\lambda-\lambda_{\rm d}$ between the incident wavelength and the nearest defect mode wavelength controls the order of the OFT. In fact, as shown in~\cite{aemebysk:apl:08,aemebysk:pra:08}, the OFT is first-order for $\delta>0$ and second-order when $\delta<0$. The ONR behavior associated with LC reorientation is summarized in Fig.~2 where the phase jump of the transmitted light field, $\delta \varphi$, and the output Poynting vector amplitude normalized to its threshold value, $\widetilde{S}_z^{\rm out} = \widetilde{S}_z^{\rm out}/\widetilde{S}_{z {\rm , th}}^{\rm in}$ are represented as a function of $\widetilde{S}_z^{\rm in}=\mathrm{sign}(k_z)\rho$, where ${\bf k}=\pm(2\pi/\lambda){\bf e}_z$ is the wavevector of the incident light, for positive [see Fig.~2(a,b)] and negative  [see  Fig.~2(c,d)] detuning values in the vicinity of the two adjacent defect modes $\lambda_{\rm d} \simeq 604$nm and $\lambda_{\rm d} \simeq 596$nm, [see Fig.~1(f)]. Note that $\widetilde{S}_z^{\rm in}>0$ for the light propagating in LR direction, and $\widetilde{S}_z^{\rm in}<0$ for opposite direction (RL). In Fig.~2, the solid and dashed curves refer to wavelength belonging to region A and B, respectively [see Fig.~1(f)].

We find a reversible ONR response both for the positive and negative detunings. However, the case $\delta>0$ looks more attractive for both phase [see Fig.~2(a)] and amplitude [see Fig.~2(b)] dependences. Indeed, $\delta \varphi$ typically experiences a $\pi$-jump during LC reorientation. This is due to the fact that the defect mode resonance shifts toward larger wavelength when the OFT occurs, hence the fixed excitation wavelength $\lambda$ passes from one to the other side of the defect mode resonance during transition form the `off' to the `on' state, i.e. the effective detuning changes it sign, which is accompanied by a $\pi$ phase shift. The amplitude also exhibits an interesting behavior. Indeed, although the overall transmission is rather small [of the order of a few percent, see Fig.~2(b)], a very large LR/RL transmission contrast can be achieved following the pump-assisting scheme proposed in~\cite{pumpassisting_OE2008}. In fact, a enhancement factor $\sim 10^3$ times larger than that obtained in~\cite{pumpassisting_OE2008} can readily be achieved in our case, which is due to the almost 100\% relative hysteresis width~\cite{aemebysk:apl:08}.

In conclusion, we have proposed a simple approach to achieve a reversible nonreciprocal optical response controlled by the wavelength of light. This became possible by using a linear/nonlinear (here made of liquid crystals) defect pair embedded in a periodic structure that is invariant by mirror symmetry when the nonlinearity is not activated. The present concept can be generalized to 2D and 3D photonic crystals geometries.

The work has been supported by the Australian Research Council.

\end{document}